\documentstyle[12pt,epsf,axodraw]{article}

\def\mypagenumber{1}

\def\myend{\end{document}}

\def\Journal#1#2#3#4{{#1}{\bf #2} (#4) #3}

\def\NPB{{\em Nucl.\ Phys.} B}
\def\PLB{{\em Phys.\ Lett.} B}

\def\PRD{{\em Phys.\ Rev.} D}

\def\RMP{{\em Rev.\ Mod.\ Phys.} }

\def\PR{{\em Phys.\ Rep.}}

\normalsize

\newcounter{sxn}

\newcounter{axn}

\date{}

\newdimen\mybaselineskip
\newcommand{\beeq}{\begin{equation}}
\newcommand{\eneq}{\end{equation}}
\newcommand{\be}{\begin{eqnarray}}
\newcommand{\ee}{\end{eqnarray}}
\newcommand{\bpic}{\begin{picture}}
\newcommand{\epic}{\end{picture}}

\def\dd{\partial}
\def\la{\raise.16ex\hbox{$\langle$} \, }
\def\ra{\, \raise.16ex\hbox{$\rangle$} }

\def\psibar{ \psi \kern-.65em\raise.6em\hbox{$-$} }
\def\mbar{ m \kern-.78em\raise.4em\hbox{$-$}\lower.4em\hbox{} }

\def\L{ {\cal L} }

\def\n@space{\nulldelimiterspace=0pt \mathsurround=0pt }
\def\huge#1{{\hbox{$\left#1\vbox to 20.5pt{}\right.\n@space$}}}

\def\myskip{\noalign{\kern 8pt}}
\def\myeqspace{\noalign{\kern 10pt}}

\def\boxit#1{$\vcenter{\hrule\hbox{\vrule\kern3pt
    \vbox{\kern3pt\hbox{#1}\kern3pt}\kern3pt\vrule}\hrule}$}
\def\bigbox#1{$\vcenter{\hrule\hbox{\vrule\kern5pt
     \vbox{\kern5pt\hbox{#1}\kern5pt}\kern5pt\vrule}\hrule}$}

\def\ignore#1{{}}

\begin{document}

\bibliographystyle{unsrt}
\footskip 1.0cm

\thispagestyle{empty}
\setcounter{page}{\mypagenumber}

%{\baselineskip=10pt \parindent=0pt \small
%\mydate 
%}                             
\begin{flushright}{\sc  TPI--MINN--97/30--T} \\{\sc    
NUC--MINN--97/14--T} \\
{\sc    HEP--MINN--97--1614} \\{\sc LBNL 41103}\\\end{flushright}
\vspace{1in}
\begin{center}{\Large \bf { Temperature Dependence of $\eta $ and 
$\eta^{\prime}$ Masses }}\\
\vspace{1in}
{\large  Jamal Jalilian-Marian$^1$ and Bayram Tekin$^{2}$}\\
\vspace{.2in}
$^1${\it  Nuclear Science Division, Lawrence Berkeley National Laboratory, 
          Berkeley, CA 94720}\\
$^2${\it School of Physics and Astronomy, University of Minnesota, 
         Minneapolis, MN 55455}\\ 
\end{center}

\vspace*{25mm}
%\baselinestretch{2.0}

%\normalsize

\begin{abstract}
\baselineskip=18pt
We investigate the temperature dependence of $\eta$ and $\eta^{\prime}$ 
masses due to scatterings from thermal pions in a heat bath using 
the non-linear sigma model. We show that mass shifts of $\eta$ and 
$\eta^{\prime}$ and the shift in the mixing angle are negligible. 
\end{abstract}

\vspace*{5mm}

%\end{titlepage}
 
\newpage

%\setcounter{page}{1}

%\textheight=20cm
%\headsep=0.75cm
%\vsize=20cm

%%%%%%%%%%%%%%%%%%%%%%%%%%%%%%%%%%%%%%%%%%%%%%%%%%%%%%%%%%%%%%%%%
\normalsize
\baselineskip=22pt plus 1pt minus 1pt
\parindent=25pt
%\vspace*{5mm}

\section{Introduction}

Temperature dependence of $\eta$ and $\eta^{\prime}$
masses is a very interesting subject and has been a hot topic 
lately~\cite{KKM,HW,TS}. The $\eta^{\prime}$,
being heavier than the other pseudoscalar mesons, naturally begs the
question  why it is so heavy. Besides being of academic interest,
this also has important experimental consequences: for instance, in
dilepton spectra which may help with identifying signals of a possible
quark-gluon plasma. 

Even though one
can understand the origin of this mass difference between the $\eta^{\prime}$
and the other pseudoscalars at $T=0$ in terms of the $U(1)$ anomaly
and instantons, a satisfactory understanding of this problem at finite
temperatures is still missing. On one hand, one might expect that at
high enough temperatures instanton effects and the anomaly will go away
and as a result $\eta^{\prime}$ will become mass degenerate with 
$\eta$ and other pseudoscalar mesons~\cite{KKM}. The first 
question that comes to mind is how high does the temperature have to be
since the relavent temperature range of $100 - 200 $ MeV may 
not be high enough in order to ignore instanton effects. On the other
hand, it has been argued that instantons do not go away in the 
temperature range of interest but rather are rearranged so that the
effect of the $U(1)$ anomaly is essentially unchanged. This leads to
the non-strange meson being heavier than the strange one if the
anomaly is strong enough~\cite{TS}.  

In this work, our goal is rather modest and pedagogical. We propose 
to use the non-linear sigma model, with  symmetry breaking and an
anomaly term added, to investigate the temperature dependence of
$\eta$ and $\eta^{\prime}$ masses for temperatures up to $150$ MeV.
We consider the change in masses of $\eta$ and $\eta^{\prime}$ due to
scatterings from a heat bath consisting mainly of pions. By evaluating
the relevant one loop diagrams at finite temperature, one can calculate
the temperature dependence of this change in masses.

One may question the validity of using the non-linear sigma model
for temperatures around the chiral phase transition temperature
$T\sim 150-200$. However, we are rather interested in gross features 
of the model with respect to the mass shifts below the phase transition
temperature and certainly do not claim to address the problem of
mass change across the phase transition temperature. As a first
approximation, this will tell us the direction and magnitude of mass
shifts.

Furthermore, we will neglect the effect of scatterings from other particles
in the heat bath like kaons, etc. since they are much heavier than pions
and one does  not expect to have too many
of them in the heat bath in temperature ranges we are interested in.
We will take the pion mass to be constant. This is certainly a good 
approximation for temperatures up to $100$ MeV but will start changing
beyond that where scattering from other particles may also become important.
This is all ignored. 

Another problem, and perhaps the central one is the temperature dependence
of the coefficient of the anomaly~\cite{GPY,DP,SH,PIS,KA,K}. The 
temperature dependence
of this coefficient is not well known and is usually modeled in the 
literature such that it goes to zero at sufficiently high temperatures.  
Here, we will take this coefficient to be constant. This is plausible
in the temperature region we are interested as can be seen, for instance,
in Fig. (1) of the paper by Sch\"{a}fer~\cite{TS} where it is shown that the
t'Hooft 
operator varies only slowly up to $T\sim 100$ MeV and that only around
$T\sim 150$ MeV  does it have a sharp increase. Again, we emphasize
that it is not our goal in this work to investigate the behavior of
masses around the chiral phase transition temperature.

We then consider the one loop diagrams with pions in the loop. These
diagrams, evaluated at finite $T$, correspond to scattering of $\eta$
or $\eta^{\prime}$ from thermal pions in the heat bath. We show that
the changes in masses are negligible. We then consider the possibility
that $\eta$ or $\eta^{\prime}$ can fluctuate into a vector or 
axial vector meson, scatter from thermal pions in the bath and then go 
back to its original state. We show that these processes are absent
and do not contribute to mass shifts of $\eta$ and $\eta^{\prime}$.

\section{Pseudoscalars} 

We will start with the following effective Lagrangian describing the
pseudoscalar nonet,

\be
\L&=& {1\over 8} F_\pi^2 \mbox{tr} \dd_\mu U \dd^\mu U^{\dagger} + 
{1\over 8} 
F_\pi^2 \mbox{tr} M (U + U^{\dagger} -2) +a (\mbox{det} U + \mbox{det} 
U^{\dagger} -2 ), \label{eq:Leff}
\ee
where $U= \mbox{exp} \left({2i\phi\over F_\pi}\right)$ and $F_{\pi}= 
\sqrt{2} f_{\pi}= 
132$ MeV is the pion decay constant. The mass matrix for the nonet is 
\be
M =
\left( \begin{array}{ccc}
m_\pi^2 &0&0 \\
0&m_\pi^2&0 \\
0&0& 2m_K^2 - m_\pi^2
\end{array}\right)
\label{eq:phimass}
\ee
%\vspace{.4in}
The first two terms in our effective Lagrangian are the usual non-linear
sigma model with massive psuedoscalar fields $\phi$ given by 
\be
\phi =
\left( \begin{array}{ccc}
{1\over{\sqrt 2}}(\pi^0 +{{\eta_8 +\sqrt 2 \eta_0}\over {\sqrt 3}}) & 
\pi^+ & K^+ \\
\pi^- & {1\over{\sqrt 2}}(-\pi^0 +{{\eta_8 +\sqrt 2 \eta_0}\over {\sqrt 
3}}) & K^0 \\
K^- & {\bar{K}}^0 & {1\over{\sqrt 2}} ({{ -2 \eta_8 +\sqrt 2 \eta_0}
\over{\sqrt 3}})
\end{array}\right)
\label{eq:psn}
\ee
%\vspace{.4in}
We have included a determinantal term in our effective Lagrangian
to account for the axial $U(1)$ anomaly. The coefficient of the  anomaly
term, $a$, is to be determined from experimental data at zero temperature.
With this effective Lagrangian at hand, we will expand the field $U$ 
and look for interaction terms between $\eta$ and $\eta^{\prime}$ and
pions. We ignore kaons altogether because they are much
heavier than pions. In other words, our heat bath consists mainly of pions.

The quadratic piece of the Lagrangian then becomes
\be
\L_{\mbox{quad}} &=& {1\over2} (\dd_\mu \eta_0)^2 +
                   {1\over2} (\dd_\mu \eta_8)^2 \nonumber \\
               &-&  {1\over 6} \left\{(- m_\pi^2 +4 m_K^2) \eta_8^2 
                    + (m_\pi^2 +2 m_K^2 + {72 a \over F_\pi^2})\eta_0^2 
                 +4\sqrt 2(m_\pi^2 - m_K^2)\eta_0 \eta_8 \right\}
\label{eq:exquad}
\ee
whereas the interaction terms are
\be
\L_{\mbox{int}}= {m_\pi^2\over 6F_\pi^2} (2\pi^+\pi^- +\pi_0^2)(\eta_8 
+\sqrt 2\eta_0)^2
\label{eq:exinter}
\ee
%\be
%tr M\phi^2 = {1\over 3} \left\{ (- m_\pi^2 +4 m_K^2)\eta_8^2 
% + 4\sqrt2(m_\pi^2 - m_K^2)\eta_8 \eta_0 + (m_\pi^2 +2 m_K^2)\eta_0^2 
%\right\} 
%\ee
%\be
%tr M\phi^4 = m_\pi^2 (2\pi^+\pi^- +\pi_0^2)(\eta_8 +\sqrt 2\eta_0)^2
%\ee
%\be
%det U + det U^{\dagger} -2  =-{12\over F_\pi^2}\eta_0^2
%\ee
It is interesting to note that there are no derivative coupling 
involving $\eta$ , $\eta^{\prime}$ with pions.

At the classical (tree) level, it is clear from the quadratic part 
of our Lagrangian that we need to
rotate the $\eta_0$ and $\eta_8$ fields and write them in terms
of the physical fields $\eta$ and $\eta^{\prime}$ defined as
\be
\eta &=& -\eta_0 \sin \theta + \eta_8 \cos \theta ,\nonumber \\
\eta^{\prime} &=& \eta_0 \cos \theta + \eta_8 \sin \theta .
\label{eq:physetas}
\ee
The mixing angle $\theta$ is defined as
\be
\tan 2\theta = {C\over {A-B}}\nonumber
\ee
where
\be
A&=&   m_\pi^2 +2 m_K^2 + {72 a \over F_\pi^2}, \nonumber \\
B&=&   -m_\pi^2 +4 m_K^2 ,\nonumber \\
C&=&  4\sqrt 2(m_\pi^2 - m_K^2) .
\ee
The physical masses of $\eta$ and $\eta^{\prime}$ are given by
\be
m^2_{\eta} &=& {1\over 3} \bigg[ B \cos^2\theta +A \sin^2\theta - 
{C\over 2}\sin 2\theta\bigg]  \nonumber \\
m^2_{\eta^{\prime}} &=& {1\over 3} \bigg[ B \cos^2\theta +A \sin^2\theta + 
{C\over 2}\sin 2\theta\bigg] .  
\ee
We determine the mixing angle and the coefficient of the anomaly by 
matching the  $\eta $ and $\eta^{\prime}$ masses with the experimental 
values. Using $m_{\pi}= 135$ MeV, $m_{K}= 497$ MeV and 
$a= {F_{\pi}^{2}\over 24} [m^2_{\eta^{\prime}} + m^2_{\eta} - 2m^2_{K}]$, 
the mixing angle comes out to be $\theta = -18.5^{\circ}$~\cite{HW}
 in close
agreement with the experimental value of $-20^{\circ}$.
This redefinition would then diagonalize the quadratic part of the Lagrangian
in the usual manner. However, here we will be considering quantum
corrections  and have to include
the quantum loop effects coming from the pion tadpole before 
diagonalizing the mass matrix.

In order to calculate the mass shifts due to scattering with thermal pions, 
all we need to do now is to construct the lowest order Feynman diagrams
contributing to the self energy  matrix $\prod$. The components of this 
matrix are  represented  diagrammatically by

\vspace{.5in}
\begin{center}
\bpic(70,12)
\put(-65,-13){$\prod_{\eta_0 \eta_0} = -2$}
\put(30,3){\circle{26}}
\put(0,-7){$\eta_0$}
\put(60,-7){$\eta_0$}
\put(30,20){$\pi_0$}
\Line(-6,-11)(66,-11)
\epic
\end{center}
%\vskip 0.5 cm
\vspace{1 cm}

\begin{center}
\bpic(70,12)
\put(-65,-13){$\prod_{\eta_8 \eta_8} = -2$}
\put(30,3){\circle{26}}
\put(0,-7){$\eta_8 $}
\put(60,-7){$\eta_8 $}
\put(30,20){$\pi_0$}
\Line(-6,-11)(66,-11)
\epic
\end{center}
\vskip .5 cm

\begin{center}
\bpic(70,12)
\put(-65,-13){$\prod_{\eta_8 \eta_0} = -$}
\put(30,3){\circle{26}}
\put(0,-7){$\eta_8 $}
\put(60,-7){$\eta_0 $}
\put(30,20){$\pi_0$}
\Line(-6,-11)(66,-11)
\put(-100, -40){Diagram 1: Relevant tadpoles for the self energy matrix.}
\epic
\end{center}
\vskip 1.5 cm 
These diagrams give
\be
\prod = -2 \lambda_{(1,2,3)} T\sum_{n} \int 
{d^3 p \over (2\pi)^3}{1 \over w_n^2 + \vec{p}^2 + m^2_{\pi}}
\label{eq:selfenergy}
\ee
where 
\be
\lambda_1 = {m_{\pi}^{2} \over 3F_{\pi}^{2}} \hskip 1 cm
\lambda_2 = {m_{\pi}^{2} \over 6F_{\pi}^{2}} \hskip 1 cm 
\lambda_3 = \sqrt 2{m_{\pi}^{2} \over 6F_{\pi}^{2}}
\ee

Since we
are interested in the temperature dependence of these corrections, we will
neglect the vacuum part of the self energy diagram and assume that it
is  renormalized as in zero temperature field theory~\cite{JK}. The 
$T$ dependent
pieces of the self energy diagram will then give the change in the masses
of $\eta_0$ and $\eta_8$. They are proportional to the integral
\be
I \equiv \int {d^3 p \over (2\pi)^3}{1 \over w}
{1 \over e^{\beta w} -1}
\label{eq:integral}
\ee
where $\beta = 1/T$ and $w= \sqrt{\vec{p}^2 + m^2_{\pi}}$. This integral
can be evaluated analytically in terms of an infinite sum of modified Bessel
functions $K_{1}(x)$\cite{ARF} with the result 

\be
I = {m_{\pi}T \over 2\pi^2} 
\sum^{\infty}_{n=1} {1 \over n} K_{1}({nm_{\pi} \over T}).
\label{eq:integralseries}
\ee
We are interested in not too high temperatures so that we will take 
$m_{\pi}= 135$ MeV to be the upper limit for $T$. 
One can then approximate the exact series
by the asymptotic limit~\cite{ARF} where, for $x > 1$,
\be
K_{\nu}(x) = \sqrt{{\pi \over 2x}}e^{-x}\bigg[1+ {(4\nu^2 -1^2)\over
1!8x} + {(4\nu^2 -1^2)(4\nu^2 -3^2)\over 2!(8x)^2} + \cdots \bigg]
\label{eq:ASSYMP}
\ee
One can check numerically that
it is enough to keep the first three terms in
the infinite series. 
It turns out to be a good approximation even when $x\sim 1$ as will be 
shown below. Keeping the first three terms, we have 
\be
I &=& - {T \over {2\pi}} \sqrt{{m_{\pi}T\over 2\pi}} 
e^{-{m_{\pi} \over T}}\bigg[ 1 +  {3T \over 8m_{\pi}} - 
{15T^2 \over 128 m_{\pi}^2} + 
 {1 \over 2\sqrt{2}} e^{-{m_{\pi}\over T}}     
 (1 + {3 T\over 16 m_\pi} - {15 T^2 \over 512 m_\pi^2}) 
\nonumber \\ &+&{1 \over 3\sqrt{3}} e^{-{2 m_{\pi}\over T}}
 (1 + {3 T\over 24m_\pi} - {15 T^2\over 1152 m_\pi^2} )\bigg] 
\label{eq:result}
\ee

To check the validity of this approximation near $x=1$ where our 
approximation would have the largest error , one can 
evaluate the value of integral $I$ exactly at this value. One obtains
$I =  0.035 m_\pi^2 $.
Equation ~(\ref{eq:result}) differs from this value by $2$ percent.  

Including these corrections in the original Lagrangian~(\ref{eq:exquad})
we get
\be
\L_{\mbox{quad}} &=& {1\over2} (\dd_\mu \eta_0)^2 +
                   {1\over2} (\dd_\mu \eta_8)^2 
          -  {1\over 6} \Bigg\{\bigg[(-1 + {2I \over F_{\pi}^{2}}) m_\pi^2 
               +4 m_K^2\bigg] \eta_8^2 \nonumber\\
          &+& \bigg[(1+ {4I \over F_{\pi}^{2}})m_\pi^2 +2 m_K^2 + 
                   {72 a \over F_\pi^2}\bigg]\eta_0^2 
          + 4\sqrt 2\bigg[(1+ {I \over {4 F_{\pi}^{2}}})m_\pi^2 - 
                 m_K^2\bigg]\eta_0 \eta_8 \Bigg\}
\label{eq:newexquad}
\ee
Now we have to diagonalize the mass matrix in order to get the $\eta$ and
$\eta^{\prime}$ masses. Define the physical fields $\eta$ and
$\eta^{\prime}$ in terms of $\eta_0$ and $\eta_8$ as
\be
\eta &=& -\eta_0 \sin \phi + \eta_8 \cos \phi \nonumber \\
\eta^{\prime} &=& \eta_0 \cos \phi + \eta_8 \sin \phi.
\label{eq:newphysetas}
\ee
The new mixing angle $\phi$, defined as 
\be
\tan 2\phi = {\tilde{C}\over{\tilde{A}-\tilde{B}}}, \nonumber
\ee
where the constants $\tilde{A},\tilde{B},\tilde{C}$ are now
\be
\tilde{A}&=&  A - {4 m_\pi^2\over F_\pi^2}I \nonumber\\
\tilde{B}&=&  B - {2 m_{\pi}^2\over F_{\pi}^{2}}I \nonumber \\
\tilde{C}&=&  C - {\sqrt2  m_{\pi}^2\over F_{\pi}^{2}}I .
\ee
The mixing angle at finite temperature is not significantly different 
from its value at zero temperature. 

At finite temperature we obtain the following formulae for the masses:
\be
m^2_{\eta}(T) &=& m^2_{\eta}(0) - { m^2_{\pi}\over 3 F^2_{\pi}}(1+ 
\sin^2\phi - {1\over 2\sqrt 2} \sin 2\phi )I \nonumber \\
m^2_{\eta^{\prime}}(T) &=& m^2_{\eta^{\prime}}(0) - { m^2_{\pi}\over 3 
F^2_{\pi}}(1+ \cos^2\phi + {1\over 2\sqrt 2} \sin 2\phi )I .
\ee
It is interesting to note that the correction to $\eta$ mass  is 
larger than the correction to mass of $\eta^{\prime}$. Since the 
integral $I$ is very
small even at $135$ MeV it is clear that neither  mass
receives a significant change. In Figures 1 and 2, we plot the 
values of $\eta$ and $\eta^{\prime}$ mass as a function of temperature.

\begin{figure}
\begin{center}
\leavevmode
\epsfysize=3.5in \epsfbox{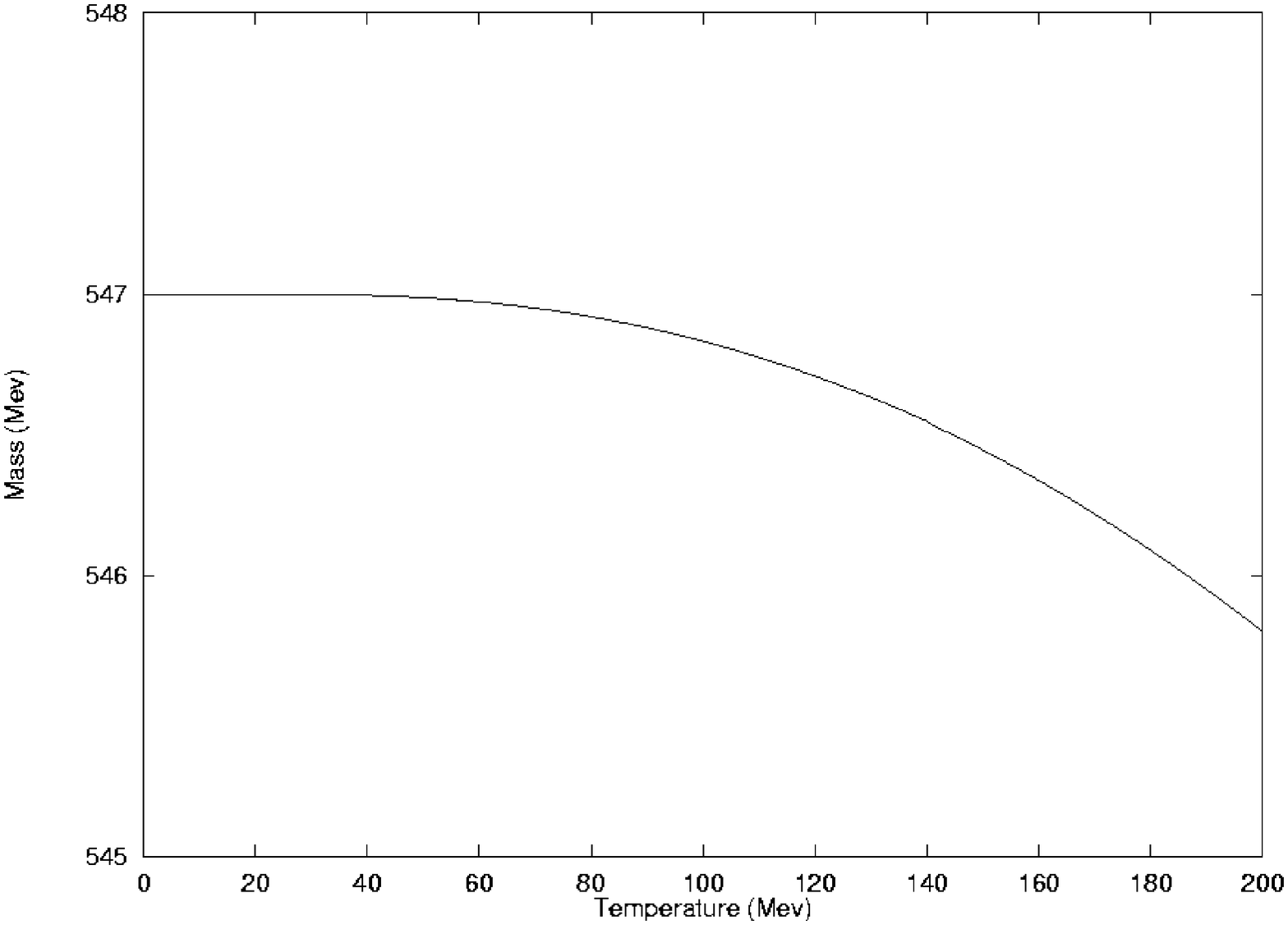}
\caption{ $m_\eta$ as a function of temperature.} 
\end{center}
\end{figure}

\begin{figure}
\begin{center}
\leavevmode
\epsfysize=3.5in \epsfbox{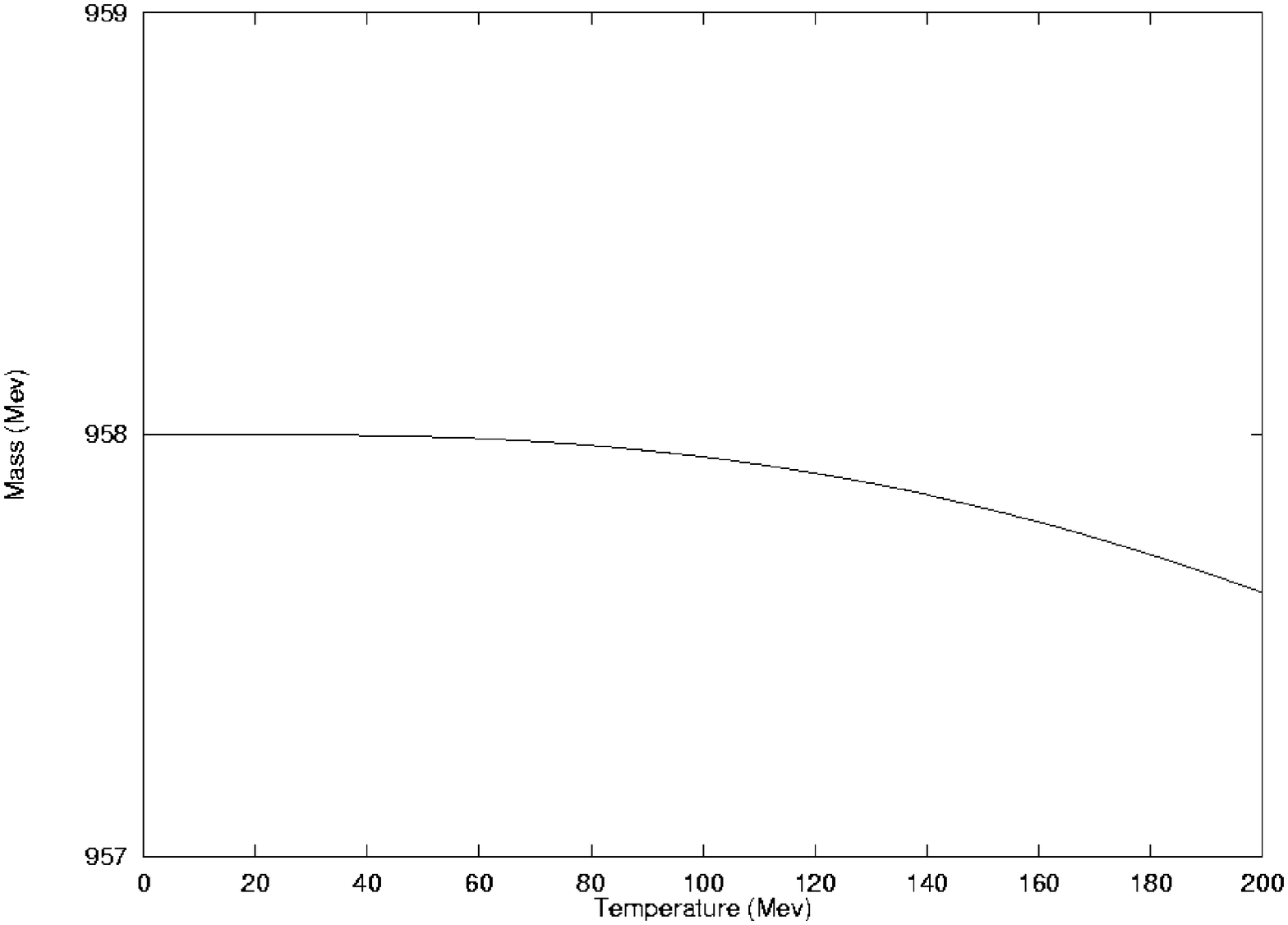}
\caption{$m_{\eta^{\prime}}$ as a function of temperature.} 
\end{center}
\end{figure}

One may naturally wonder if inclusion of vector and axial vector mesons
would change our results drastically. For instance, interaction of the 
type depicted in Diagram (2) can contribute to the change in the $\eta$, 
$\eta^{\prime}$ masses.

\vspace{.5in}
\begin{center}
\bpic(70,12)
\put(-65,1){$\prod_{\eta} = -2$}
\put(50,3){\circle{26}}
\put(20,-7){$\eta$}
\put(80,-7){$\eta$}
\put(47,20){$\pi_0$}
\put(45,-22){V,A}
\Line(14,3)(37,3)
\Line(63,3)(86,3)
\put (-100,- 40){ Diagram 2: Vector/axial vector contribution. }
\epic
\end{center}
\vskip 0.5 cm
\vspace{.5in}
To investigate such a possibility, we will need to write an effective
Lagrangian which includes both vectors and axial vectors in addition
to our pseudoscalar nonet. We will start with
the following effective Lagrangian introduced by Meissner\cite{MEIS}
based on massive Yang-Mills gauge theory

\be
 \L =\L_0 +\L_1 + \L_2 + \L_3 \nonumber
\ee  
where 
\be
\L_0 &=& {1\over 8 }\mbox{tr}[(D_\mu U) (D_\mu U)^\dagger] ,\nonumber \\
\L_1 &=& -{1\over 2} \mbox{tr}[F^L_{\mu\nu}F^{\mu\nu L} 
+F^R_{\mu\nu}F^{\mu\nu R}]
       +\gamma \mbox{tr} [F^L_{\mu\nu}UF^{\mu\nu R}U^\dagger] ,\nonumber \\
\L_2 &=& m_0^2 \mbox{tr}[A^L_\mu A^{\mu L} +A^R_\mu A^{\mu R}] + 
         B \mbox{tr}[A^L_\mu U A^{\mu R}U^\dagger], \nonumber \\
\L_3 &=& {1\over 8}
        F_\pi^2 \mbox{tr} M (U + U^{\dagger} -2) +a (\mbox{det} U + 
\mbox{det} U^{\dagger} -2 ).      
\ee
The covariant derivative is defined as
\be
D_{\mu} U = \dd_{\mu} -igA_{\mu}^{L} + ig A^{R}_{\mu}
\ee
where the field strength is
\be
F^{L,R}_{\mu \nu} = \dd_{\mu} A^{L,R}_{\nu} - \dd_{\nu}A^{L,R}_{\mu} 
-ig [A^{L,R}_{\mu},A^{L,R}_{\nu}] 
\ee
and left-handed and right-handed fields $A^{L,R}_{\mu}$ are 
defined as
\be
A^{L}_{\mu} = {1 \over 2} (V_{\mu} + A_{\mu}),\,\,\,\,\,\,
A^{R}_{\mu} = {1 \over 2} (V_{\mu} - A_{\mu}) .
\ee
Notice that at this point both vector and axial vector mesons have the
same mass $m_{0}^{2}$ and $\gamma$ and $B$ are related to 
$m_{\rho}$~\cite{MEIS}.

In order to avoid the difficulties related to experimental uncertainties
with regard to axial vector mesons, one can eliminate the explicit 
dependence of the Lagrangian on axial vectors by a suitable gauge
transformations such that all axial vectors are zero in this new gauge
as done, for example, in \cite{MEIS}. We then have
\be
\L_1 &=& (\gamma -1 ) \mbox{tr}[F_{\mu 
\nu}(\rho)F^{\mu\nu}(\rho)] ,\nonumber\\
\L_2 &=& (B + 2 m_0^2)\mbox{tr}[\rho_\mu \rho^\mu] +(i/g)(B +2 m_0^2)
         \mbox{tr}[\rho^\mu (\dd_\mu U^{1/2} U^{-1/2} 
         + \dd_\mu U^{-1/2} U^{1/2}] \nonumber \\
     &+ & 2(m_0^2/g^2) \mbox{tr}[\dd_\mu U^{1/2}\dd^\mu U^{-1/2}]
         - (B/g^2)\mbox{tr}[U^{-1/2}\dd_\mu U^{-1/2} U^{1/2}\dd^\mu U^{1/2}]
\ee          
Choosing $\gamma = 3/4$ will give the correct kinetic term for the vector
mesons. Also, $m_{v}^{2}= 4m_{0}^{2} + 2B$. For a discussion of these
parameters see \cite{MEIS}.
In order to get the appropriate interaction terms for pseudoscalars and
vector mesons we expand the Lagrangian in terms of the pseudoscalar
fields as before and examine the relevant vertex, which is of the form
\be
\mbox{tr} V_{\mu}\phi\stackrel{\leftrightarrow}{\partial}_{\mu}\phi =
V_{\mu}^{a} \phi^{b}\partial_{\mu} \phi^{c} 
\mbox{tr} \lambda^{a} 
[\lambda^b,\lambda^c]
\label{eq:vectorphiphi}
\ee
where  $\lambda^0 = {\bf 1}$ while for $a=1,\cdots 8$ they are the usual
Gell-Mann matrices. Since we are interested in interactions of $\eta_0$
and $\eta_8$ (corresponding to $\lambda^0$ and $\lambda^8$) and 
pions ($\lambda^{1,2,3}$) with vectors, then $a,b,c = 0,1,2,3,8$ only.
It is then easy to show that this term does not
involve interaction vertices of the type $\eta \pi \rho$
which would contribute to the change in masses of $\eta$ and 
$\eta^{\prime}$ through diagrams depicted in Diagram (2). 

\section{Conclusion}

We used the non-linear sigma model, with appropriately added symmetry 
breaking and anomaly terms, to investigate the shifts in
$\eta$ and $\eta^{\prime}$ masses due to scatterings with thermal
pions in a heat bath. We showed that the mass shifts as well as the change
in the mixing angle are negligible. We 
also considered possible effects of vector and axial vector mesons
on $\eta$ and $\eta^{\prime}$ masses and showed that vector and axial
vector mesons do not contribute. Our straight-forward calcualtion imply 
that if experiments show a stronger 
temperature dependence , then one would conclude that non-linear sigma
model description fails to be true and one should look for new physics.  

\leftline{\bf Acknowledgements} We would like to thank Joe Kapusta
for suggesting the problem and for his critical reading of the manuscript.
We would also like to thank V. Koch, T. Sch\"{a}fer, J. Schaffner-Bielich
for useful discussions
and Ismail G\"{u}ler for his help with the figures. This work was
supported by the Director, Office of Energy Research, Office of High Energy
and Nuclear Physics Division of the Deparment of Energy, under 
contract No. DE-AC03-76SF00098 and DE-FG02-87ER40328.

\leftline{\bf References}

\renewenvironment{thebibliography}[1]
        {\begin{list}{[$\,$\arabic{enumi}$\,$]}  % {\arabic{enumi}.}
        {\usecounter{enumi}\setlength{\parsep}{0pt}
         \setlength{\itemsep}{0pt}  \renewcommand{\baselinestretch}{1.2}
         \settowidth
        {\labelwidth}{#1 ~ ~}\sloppy}}{\end{list}}


\begin{thebibliography}{99}

\small

\bibitem{KKM}
J.\ Kapusta,  D.\ Kharzeev and L.\ McLerran, \Journal{\PRD}{53}{1996}{5028}. 

\bibitem{HW}
Z.\ Huang and X-N.\ Wang, \Journal{\PRD}{53}{1996}{5034}.   

\bibitem{TS}
T. Sch\"{a}ffer, \Journal{\PLB}{389}{1996}{445}. 

\bibitem{GPY}
D.J. Gross, R.D. Pisarski and L.G. Yaffe, \Journal{\RMP}{53}{1981}{43}.

\bibitem{DP}
D.I. Dyakonov and V.Y. Petrov, \Journal{\NPB}{245}{1984}{259},
{272}{1986}{475}.

\bibitem{SH}
E. Shuryak, \Journal{\NPB}{302}{1988}{559}.

\bibitem{PIS}
R.D. Pisarski and L.G. Yaffe, \Journal{\PLB}{97}{1980}{110}.

\bibitem{KA}
H. Kikuchi and T. Akiba, \Journal{\PLB}{200}{1988}{543}.

\bibitem{K}
T. Kunihiro, \Journal{\NPB}{351}{1991}{593}.

\bibitem{JK}
J. I. Kapusta, {\it Finite Temperature Field Theory}, Cambridge 
University Press 1993.

\bibitem{ARF}
G. Arfken, {\it Mathematical Methods for Physicists}, Academic Press 1985.

\bibitem{MEIS}
U.G.\ Meissner, \Journal{\PR}{161}{1987}{213}.

\end{thebibliography}
\end{document}